\documentclass[11pt]{elsart}

\usepackage{amssymb,amsmath}
\usepackage{mathrsfs} 
\usepackage{graphicx}

\begin{document}

\begin{frontmatter}
\title{Two-agent Nash implementation: \\A new result}
\author{Haoyang Wu\corauthref{cor}}
\corauth[cor]{Wan-Dou-Miao Research Lab, Shanghai, 200051, China.}
\ead{hywch@mail.xjtu.edu.cn} \ead{Tel: 86-18621753457}

\begin{abstract}

[Moore and Repullo, \emph{Econometrica} \textbf{58} (1990)
1083-1099] and [Dutta and Sen, \emph{Rev. Econom. Stud.} \textbf{58}
(1991) 121-128] are two fundamental papers on two-agent Nash
implementation. Both of them are based on Maskin's classic paper
[Maskin, \emph{Rev. Econom. Stud.} \textbf{66} (1999) 23-38]. A
recent work [Wu, http://arxiv.org/abs/1002.4294, \emph{Inter. J.
Quantum Information}, 2010 (accepted)] shows that when an additional
condition is satisfied, the Maskin's theorem will no longer hold by
using a quantum mechanism. Furthermore, this result holds in the
macro world by using an algorithmic mechanism. In this paper, we
will investigate two-agent Nash implementation by virtue of the
algorithmic mechanism. The main result is: The sufficient and
necessary conditions for Nash implementation with two agents shall
be amended, not only in the quantum world, but also in the macro
world.
\end{abstract}


\end{frontmatter}

\section{Introduction}
Game theory and mechanism design play important roles in economics.
Game theory aims to investigate rational decision making in conflict
situations, whereas mechanism design just concerns the
\emph{reverse} question: given some desirable outcomes, can we
design a game that produces them? Ref. \cite{Maskin1999} is seminal
work in the field of mechanism design. It provides an almost
complete characterization of social choice rules that are Nash
implementable when the number of agents is at least three. In 1990,
Moore and Repullo \cite{MR1990} gave a necessary and sufficient
condition for Nash implementation with two agents and many agents.
Dutta and Sen \cite{DS1991} independently gave an equivalent result
for two-agent Nash implementation. In 2009, Busetto and Codognato
\cite{BC2009} gave an amended necessary and sufficient condition for
two-agent Nash implementation. These papers together construct a
framework for two-agent Nash implementation.

In 2010, Wu \cite{Wu2010} claimed that the Maskin's theorem are
amended by virtue of a quantum mechanism, i.e., a social choice rule
that is monotonic and satisfies no-veto will not be Nash
implementable if it satisfies an additional condition. Although
current experimental technologies restrict the quantum mechanism to
be commercially available, Wu \cite{sim2011} propose an algorithmic
mechanism that amends the sufficient and necessary conditions for
Nash implementation with three or more agents in the macro world.
Inspired by these results, it is natural to ask what will happen if
the algorithmic mechanism can be generalized to two-agent Nash
implementation. This paper just concerns this question.

The rest of this paper is organized as follows: Section 2 recalls
preliminaries of two-agent Nash implementation given by Moore and
Repullo \cite{MR1990}. Section 3 and 4 are the main parts of this
paper, in which we will propose two-agent quantum and algorithmic
mechanisms respectively. Section 5 draws the conclusions. In
Appendix, we explain that the social choice rule given in Section 3
satisfies condition $\mu2$ defined by Moore and Repullo.

\section{Preliminaries}
Consider an environment with a finite set $I=\{1,2\}$ of agents, and
a (possibly infinite) set $A$ of feasible outcomes. The profile of
the agents' preferences over outcomes is indexed by
$\theta\in\Theta$, where $\Theta$ is the set of preference profiles.
Under $\theta$, agent $j\in I$ has preference ordering
$R_{j}(\theta)$ on the set $A$. Let $P_{j}(\theta)$ denote the
strict preference relation corresponding to $R_{j}(\theta)$.

For any $j\in I$, $\theta\in\Theta$ and $a\in A$, let
$L_{j}(a,\theta)$ be the lower contour set of agent $j$ at $a$ under
$\theta$, i.e., $L_{j}(a,\theta)=\{\hat{a}\in A:
aR_{j}(\theta)\hat{a}\}$. For any $j\in I$, $\theta\in\Theta$ and
$C\subseteq A$, let $M_{j}(C,\theta)$ be the set of maximal elements
in $C$ for agent $j$ under $\theta$, i.e.,
$M_{j}(C,\theta)=\{\hat{c}\in C: \hat{c}R_{j}(\theta)c\mbox{, for
all }c\in C\}$.

A \emph{social choice rule} (SCR) is a correspondence $f:
\Theta\rightarrow A$ that specifies a nonempty set
$f(\theta)\subseteq A$ for each preference profile $\theta\in
\Theta$. A \emph{mechanism} is a function $g: S\rightarrow A$ that
specifies an outcome $g(s)\in A$ for each vector of strategies
$s=(s_{1},s_{2})\in S=S_{1}\times S_{2}$, where $S_{j}$ denotes
agent $j$'s strategy set.

A mechanism $g$ together with a preference profile $\theta\in\Theta$
defines a game in normal form. Let $NE(g,\theta)\subseteq S$ denote
the set of pure strategy Nash equilibria of the game $(g,\theta)$. A
mechanism $g$ is said to \emph{Nash implement} an SCR $f$ if for all
$\theta\in\Theta$, $\{g(s): s\in NE(g,\theta)\}=f(\theta)$.

\textbf{Condition $\mu$}: There is a set $B\subseteq A$, and for
each $j\in I$, $\theta\in\Theta$, and $a\in f(\theta)$, there is a
set $C_{j}(a,\theta)\subseteq B$, with $a\in
M_{j}(C_{j}(a,\theta),\theta)$ such that for all
$\theta^{*}\in\Theta$, (i), (ii) and (iii) are satisfied: \\
(i) if $a\in M_{1}(C_{1}(a,\theta),\theta^{*})\cap
M_{2}(C_{2}(a,\theta),\theta^{*})$, then $a\in f(\theta^{*})$;\\
(ii) if $c\in M_{j}(C_{j}(a,\theta),\theta^{*})\cap
M_{k}(B,\theta^{*})$, for $j,k\in I$, $j\neq k$, then $c\in
f(\theta^{*})$;\\
(iii) if $d\in M_{1}(B,\theta^{*})\cap M_{2}(B,\theta^{*})$, then
$d\in f(\theta^{*})$;

\textbf{Condition $\mu2$}: Condition $\mu$ holds. In addition, for
each 4-tuple $(a,\theta,b,\phi)\in A\times\Theta\times A
\times\Theta$, with $a\in f(\theta)$ and $b\in f(\phi)$, there
exists $e=e(a,\theta,b,\phi)$ contained in $C_{1}(a,\theta)\cap
C_{2}(b,\phi)$ such that for all $\theta^{*}\in\Theta$, (iv) is
satisfied:\\
(iv) if $e\in M_{1}(C_{1}(a,\theta),\theta^{*})\cap
M_{2}(C_{2}(b,\phi),\theta^{*})$, then $e\in f(\theta^{*})$.

\textbf{Theorem 1 (Moore and Repullo, 1990)}: Suppose that there are
two agents. Then a social choice rule $f$ is Nash implementable if
and only if it satisfies condition $\mu2$.

To facilitate the following discussion, here we cite the
Moore-Repullo's mechanism as follows: For each agent $j\in I$, Let
$S_{j}=\{(\theta_{j},a_{j},b_{j},n_{j})\in\Theta \times A\times
B\times N: a_{j}\in f(\theta_{j})\}$, where $N$ denotes the set of
non-negative integers, and define the mechanism $g: S\rightarrow A$
such that for any $s\in S$:\\
(1) if $(a_{1}, \theta_{1})=(a_{2}, \theta_{2})=(a, \theta)$, then
$g(s)=a$;\\
(2) if $(a_{1}, \theta_{1})\neq (a_{2}, \theta_{2})$ and
$n_{1}=n_{2}=0$, then $g(s)=e(a_{2},\theta_{2},a_{1},\theta_{1})$;\\
(3) if $(a_{1}, \theta_{1})\neq (a_{2}, \theta_{2})$ and
$n_{1}>n_{2}=0$, then $g(s)=b_{1}$ if $b_{1}\in
C_{1}(a_{2},\theta_{2})$, and $g(s)=e(a_{2},
\theta_{2},a_{1},\theta_{1})$ otherwise;\\
(4) if $(a_{1}, \theta_{1})\neq (a_{2}, \theta_{2})$ and
$n_{2}>n_{1}=0$, then $g(s)=b_{2}$ if $b_{2}\in
C_{2}(a_{1},\theta_{1})$, and $g(s)=e(a_{2},
\theta_{2},a_{1},\theta_{1})$ otherwise;\\
(5) if $(a_{1}, \theta_{1})\neq (a_{2}, \theta_{2})$ and $n_{1}\geq
n_{2}>0$, then $g(s)=b_{1}$;\\
(6) if $(a_{1}, \theta_{1})\neq (a_{2}, \theta_{2})$ and
$n_{2}>n_{1}>0$, then $g(s)=b_{2}$.

\section{A two-agent quantum mechanism}
In this section, first we will show an example of a
Pareto-inefficient two-agent SCR $f$ that satisfies condition
$\mu2$, i.e., it is Nash implementable according to Moore-Repullo's
mechanism. Then, we will propose a two-agent version of quantum
mechanism, which amends the sufficient and necessary conditions for
Nash implementation for two agents. Hence, $f$ will not be Nash
implementable in the quantum domain.

\subsection{A Pareto-inefficient two-agent SCR}
Consider an SCR $f$ given in Table 1. $I=\{1, 2\}$,
$A=\{a^{1},a^{2},a^{3},a^{4}\}$, $\Theta=\{\theta^{1},\theta^{2}\}$.
In each preference profile, the preference relations over the
outcome set $A$ and the corresponding SCR $f$ are given in Table 1.
$f$ is Pareto-inefficient from the viewpoint of two agents because
in the preference profile $\theta=\theta^{2}$, both agents prefer a
Pareto-efficient outcome $a^{1}\in f(\theta^{1})$: for each agent
$j\in I$, $a^{1}P_{j}(\theta^{2})a^{2}$. However, since $f$
satisfies condition $\mu2$ (see the Appendix), it is Nash
implementable according to Moore-Repullo's theorem.

\emph{Table 1. A Pareto-inefficient two-agent SCR $f$ that satisfies
condition $\mu2$.}
\begin{table}[ph]
{\begin{tabular}{cccc}
 \multicolumn{2}{c}{$\theta^{1}$}&\multicolumn{2}{c}{$\theta^{2}$}\\
 $\mbox{agent }1$&$\mbox{agent }2$ &$\mbox{agent }1$&$\mbox{agent }2$ \\ \hline
 $a^{3}$&$a^{2}$ &$a^{4}$&$a^{3}$ \\
 $a^{1}$&$a^{1}$ &$a^{1}$&$a^{1}$ \\
 $a^{2}$&$a^{4}$ &$a^{2}$&$a^{2}$ \\
 $a^{4}$&$a^{3}$ &$a^{3}$&$a^{4}$ \\\hline
 \multicolumn{2}{c}{$f(\theta^{1})=\{a^{1}\}$}&\multicolumn{2}{c}{$f(\theta^{2})=\{a^{2}\}$}\\\hline
\end{tabular}}
\end{table}

\subsection{A two-agent quantum mechanism}
Following Ref. \cite{Wu2010}, here we will propose a two-agent
quantum mechanism to help agents combat ``bad'' social choice
functions. According to Eq (4) in Ref. \cite{Flitney2007},
two-parameter quantum strategies are drawn from the set:
\begin{equation}
\hat{\omega}(\theta,\phi)\equiv \begin{bmatrix}
  e^{i\phi}\cos(\theta/2) & i\sin(\theta/2)\\
  i\sin(\theta/2) & e^{-i\phi}\cos(\theta/2)
\end{bmatrix},
\end{equation}
$\hat{\Omega}\equiv\{\hat{\omega}(\theta,\phi):\theta\in[0,\pi],\phi\in[0,\pi/2]\}$,
$\hat{J}\equiv\cos(\gamma/2)\hat{I}^{\otimes
n}+i\sin(\gamma/2)\hat{\sigma_{x}}^{\otimes
  n}$,
where $\gamma$ is an entanglement measure, and
$\hat{I}\equiv\hat{\omega}(0,0)$,
$\hat{D}\equiv\hat{\omega}(\pi,\pi/2)$,
$\hat{C}\equiv\hat{\omega}(0,\pi/2)$.

Without loss of generality, we assume:\\
1) Each agent $j\in I$ has a quantum coin $j$ (qubit) and a
classical card $j$ . The basis vectors $|C\rangle\equiv(1,0)^{T}$,
$|D\rangle\equiv(0,1)^{T}$ of a quantum coin denote head up and tail
up respectively.\\
2) Each agent $j\in I$ independently performs a local unitary
operation on his/her own quantum coin. The set of agent $j$'s
operation is $\hat{\Omega}_{j}=\hat{\Omega}$. A strategic operation
chosen by agent $j$ is denoted as
$\hat{\omega}_{j}\in\hat{\Omega}_{j}$. If
$\hat{\omega}_{j}=\hat{I}$, then
$\hat{\omega}_{j}(|C\rangle)=|C\rangle$,
$\hat{\omega}_{j}(|D\rangle)=|D\rangle$; If
$\hat{\omega}_{j}=\hat{D}$, then
$\hat{\omega}_{j}(|C\rangle)=|D\rangle$,
$\hat{\omega}_{j}(|D\rangle)=|C\rangle$. $\hat{I}$ denotes
``\emph{Not flip}'', $\hat{D}$ denotes ``\emph{Flip}''. \\
3) The two sides of a card are denoted as Side 0 and Side 1. The
information written on the Side 0 (or Side 1) of card $j$ is denoted
as $card(j,0)$ (or $card(j,1)$). A typical card of agent $j$ is
described as $c_{j}=(card(j,0),card(j,1))\in S_{j}\times S_{j}$,
where $S_{j}$ is defined in Moore-Repullo's mechanism. The set
of $c_{j}$ is denoted as $C_{j}\equiv S_{j}\times S_{j}$.\\
4) There is a device that can measure the state of two quantum coins
and send strategies to the designer.

Note that if $\hat{\Omega}_{j}$ is restricted to be $\{\hat{I},
\hat{D}\}$, then $\hat{\Omega}_{j}$ is equivalent to \{\emph{Not
flip}, \emph{Flip}\}.

\textbf{Definition 1}: A \emph{two-agent quantum mechanism} is
defined as $\hat{G}:\hat{S}\rightarrow A$, where
$\hat{S}=\hat{S}_{1}\times\hat{S}_{2}$,
$\hat{S}_{j}=\hat{\Omega}_{j}\times C_{j}$ ($j\in I$). $\hat{G}$ can
also be written as
$\hat{G}:(\hat{\Omega}_{1}\otimes\hat{\Omega}_{2})\times(C_{1}\times
C_{2})\rightarrow A$, where $\otimes$ represents tensor product.

We shall use $\hat{S}_{-j}$ to express $\hat{\Omega}_{k}\times
C_{k}$ ($k\neq j$), and thus, a strategy profile is
$\hat{s}=(\hat{s}_{1},\hat{s}_{2})$, where
$\hat{s}_{1}=(\hat{\omega}_{1},c_{1})\in\hat{S}_{1}$,
$\hat{s}_{2}=(\hat{\omega}_{2},c_{2})\in\hat{S}_{2}$. A \emph{Nash
equilibrium} of $\hat{G}$ played in a preference profile $\theta$ is
a strategy profile $\hat{s}^{*}=(\hat{s}^{*}_{1},\hat{s}^{*}_{2})$
such that for any agent $j\in I$, $\hat{s}_{j}\in\hat{S}_{j}$,
$\hat{G}(\hat{s}^{*}_{1},\hat{s}^{*}_{2})R_{j}(\theta)
\hat{G}(\hat{s}_{j},\hat{s}^{*}_{-j})$. For each $\theta\in\Theta$,
the pair ($\hat{G}, \theta$) defines a game in normal form. Let
$NE(\hat{G},\theta)\subseteq \hat{S}$ denote the set of pure
strategy Nash equilibria of the game $(\hat{G},\theta)$. Fig. 1
illustrates the setup of the two-agent quantum mechanism $\hat{G}$.
Its working steps are shown as follows:

\begin{figure}[!t]
\centering
\includegraphics[height=2.3in,clip,keepaspectratio]{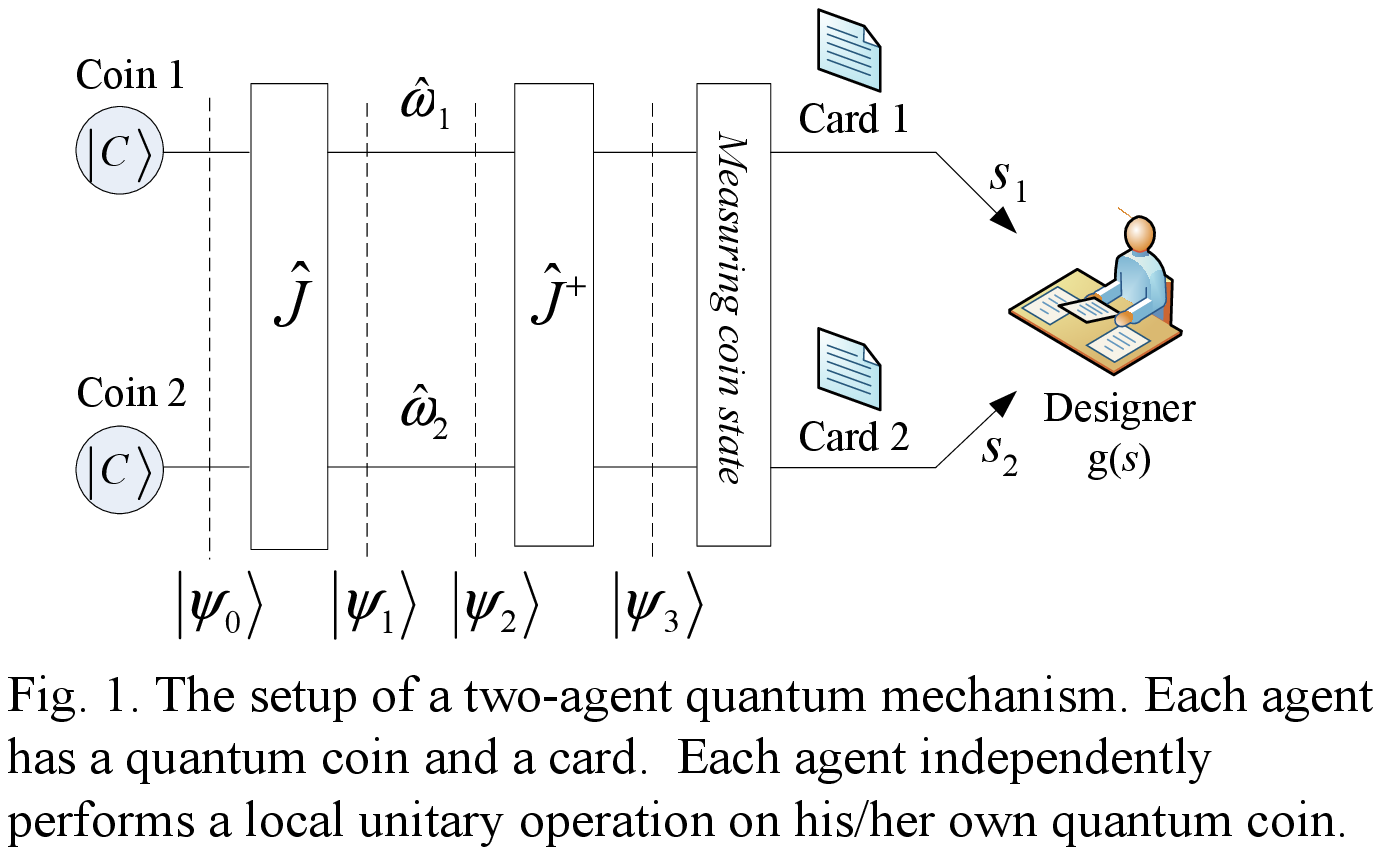}
\end{figure}

Step 1: The state of each quantum coin is set as $|C\rangle$. The
initial state of the two quantum coins is
$|\psi_{0}\rangle=|CC\rangle$.\\
Step 2: Given a preference profile $\theta$, if the two following
conditions are satisfied, goto Step 4:\\
1) There exists $\theta'\in\Theta$, $\theta'\neq \theta$ such that
$a'R_{j}(\theta)a$ (where $a'\in f(\theta')$, $a\in f(\theta)$) for
each agent $j\in I$, and $a'P_{k}(\theta)a$ for at least one agent
$k\in I$;\\
2) If there exists $\theta''\in\Theta$, $\theta''\neq \theta'$ that
satisfies the former condition, then $a'R_{j}(\theta)a''$ (where
$a'\in f(\theta')$, $a''\in f(\theta'')$) for each agent $j\in I$,
and $a'P_{k}(\theta)a''$ for at least one agent $k\in I$.\\
Step 3: Each agent $j$ sets
$c_{j}=((\theta_{j},a_{j},b_{j},n_{j}),(\theta_{j},a_{j},b_{j},n_{j}))\in
S_{j}\times S_{j}$ and
$\hat{\omega}_{j}=\hat{I}$. Goto Step 7.\\
Step 4: Each agent $j$ sets $c_{j}=((\theta',
a',*,0),(\theta_{j},a_{j},b_{j},n_{j}))$. Let the two quantum coins
be entangled by $\hat{J}$. $|\psi_{1}\rangle=\hat{J}|CC\rangle$.\\
Step 5: Each agent $j$ independently performs a local unitary
operation $\hat{\omega}_{j}$ on his/her own quantum coin.
$|\psi_{2}\rangle=[\hat{\omega}_{1}\otimes\hat{\omega}_{2}]\hat{J}|CC\rangle$.\\
Step 6: Let the two quantum coins be disentangled by $\hat{J}^{+}$.
$|\psi_{3}\rangle=\hat{J}^{+}[\hat{\omega}_{1}\otimes\hat{\omega}_{2}]
\hat{J}|CC\rangle$.\\
Step 7: The device measures the state of the two quantum coins and
sends $card(j,0)$ (or $card(j,1)$) as the strategy $s_{j}$ to the
designer if the
state of quantum coin $j$ is $|C\rangle$ (or $|D\rangle$).\\
Step 8: The designer receives the overall strategy $s=(s_{1},
s_{2})$ and let the final outcome be $g(s)$ using rules (1)-(6) of
the Moore-Repullo's mechanism. END.

Given two agents, consider the payoff to the second agent, we denote
by $\$_{CC}$ the expected payoff when the two agents both choose
$\hat{I}$ (the corresponding collapsed state is $|CC\rangle$), and
denote by $\$_{CD}$ the expected payoff when the first agent choose
$\hat{I}$ and the second agent chooses $\hat{D}$ (the corresponding
collapsed state is $|CD\rangle$). $\$_{DD}$ and $\$_{DC}$ are
defined similarly. For the case of two-agent Nash implementation,
the condition $\lambda$ in Ref. [5] is reformulated as the following
condition $\lambda'$: \\
1) $\lambda'_{1}$: Given an SCR $f$, a preference profile $\theta\in
\Theta$ and $a\in f(\theta)$, there exists $\theta'\in\Theta$,
$\theta'\neq \theta$ such that $a'R_{j}(\theta)a$ (where $a'\in
f(\theta')$, $a\in f(\theta)$) for each agent $j\in I$, and
$a'P_{k}(\theta)a$ for at least one agent $k\in I$. In going from
$\theta'$ to $\theta$ both agents encounter a preference change
around $a'$.\\
2) $\lambda'_{2}$: If there exists $\theta''\in\Theta$,
$\theta''\neq \theta'$ that satisfies $\lambda'_{1}$, then
$a'R_{j}(\theta)a''$ (where $a'\in f(\theta')$, $a''\in
f(\theta'')$) for each agent $j\in
I$, and $a'P_{k}(\theta)a''$ for at least one agent $k\in I$.\\
3) $\lambda'_{3}$: For each agent $j\in I$, let him/her be the
second agent and consider his/her payoff, $\$_{CC}>\$_{DD}$. \\
4) $\lambda'_{4}$: For each agent $j\in I$, let him/her be the
second agent and consider his/her payoff,
$\$_{CC}>\$_{CD}\cos^{2}\gamma+\$_{DC}\sin^{2}\gamma$.

\textbf{Proposition 1:} For two agents, given a preference profile
$\theta\in\Theta$ and a ``bad'' SCR $f$ (from the viewpoint of
agents) that satisfies condition $\mu2$, agents who satisfies
condition $\lambda'$ can combat the ``bad'' SCR $f$ by virtue of a
two-agent quantum mechanism $\hat{G}:\hat{S}\rightarrow A$, i.e.,
there exists a Nash equilibrium $\hat{s}^{*}\in NE(\hat{G},\theta)$
such that $\hat{G}(\hat{s}^{*})\notin f(\theta)$.

The proof is straightforward according to Proposition 2 in Ref.
\cite{Wu2010}. Let us reconsider the SCR $f$ given in Section 3.1.
Obviously, when the true preference profile is $\theta^{2}$, the two
conditions in Step 2 of $\hat{G}$ are satisfied. Hence, $\hat{G}$
will enter Step 4. In Step 4, two agents set
$c_{1}=((\theta^{1},a^{1},*,0),(\theta^{2},a^{2},*,0))$,
$c_{2}=((\theta^{1},a^{1},*,0),(\theta^{2},a^{2},*,0))$. For any
agent $j\in I$, let him/her be the second agent. Consider the payoff
of the second agent, suppose $\$_{CC}=3$ (the corresponding outcome
is $a^{1}$), $\$_{CD}=5$ (the corresponding outcome is
$e(a^{1},\theta^{1},a^{2},\theta^{2})= a^{4}$ if $j=1$, and
$e(a^{2},\theta^{2},a^{1},\theta^{1})=a^{3}$ if $j=2$), $\$_{DC}=0$
(the corresponding outcome is
$e(a^{2},\theta^{2},a^{1},\theta^{1})=a^{3}$ if $j=1$, and
$e(a^{1},\theta^{1},a^{2},\theta^{2})=a^{4}$ if $j=2$), $\$_{DD}=1$
(the corresponding outcome is $a^{2}$). Hence, condition
$\lambda'_{3}$ is satisfied, and condition $\lambda'_{4}$ becomes:
$3\geq5\cos^{2}\gamma$. If $\sin^{2}\gamma\geq0.4$, condition
$\lambda'_{4}$ is satisfied.

Therefore, in the preference profile $\theta=\theta^{2}$, there
exists a novel Nash equilibrium
$\hat{s}^{*}=(\hat{s}^{*}_{1},\hat{s}^{*}_{2})$, where
$\hat{s}^{*}_{1}=\hat{s}^{*}_{2}=(\hat{C},((\theta^{1},a^{1},*,0),(\theta^{2},a^{2},*,0)))$,
such that in Step 8 the strategy received by the designer is
$s=(s_{1},s_{2})$, where $s_{1}=s_{2}=(\theta^{1},a^{1},*,0)$.
Consequently, $\hat{G}(\hat{s}^{*})=g(s)=a^{1}\notin
f(\theta^{2})=\{a^{2}\}$, i.e., the Moore and Repullo's theorem does
not hold for the ``bad'' social choice rule $f$ by virtue of the
two-agent quantum mechanism $\hat{G}$.

\section{A two-agent algorithmic mechanism}
Following Ref. \cite{sim2011}, in this section we will propose a
two-agent algorithmic mechanism to help agents benefit from the
two-agent quantum mechanism immediately.

\subsection{Matrix representations of quantum states}
In quantum mechanics, a quantum state can be described as a vector.
For a two-level system, there are two basis vectors: $(1,0)^{T}$ and
$(0,1)^{T}$. In the beginning, we define:
\begin{align*}
|C\rangle=[1,0]^{T},\; |D\rangle=[0,1]^{T},\;
|CC\rangle=[1,0,0,0]^{T},
\end{align*}
\begin{align*}
\hat{J}=\begin{bmatrix}
  \cos(\gamma/2) & 0 &  0 & i\sin(\gamma/2)\\
  0 & \cos(\gamma/2) & i\sin(\gamma/2) & 0 \\
  0 & i\sin(\gamma/2)& \cos(\gamma/2) &  0 \\
  i\sin(\gamma/2) & 0 & 0 & \cos(\gamma/2)
\end{bmatrix},
\;\gamma\in[0,\pi/2].
\end{align*}
For $\gamma=\pi/2$,
\begin{align*}
&\hat{J}_{\pi/2}=\frac{1}{\sqrt{2}}\begin{bmatrix}
  1 & 0& 0& i\\
  0 & 1& i& 0\\
  0 & i& 1& 0\\
  i & 0& 0& 1
\end{bmatrix}.
\end{align*}
\begin{align*}
|\psi_{1}\rangle=\hat{J}|CC\rangle=\begin{bmatrix}
  \cos(\gamma/2)\\
  0\\
  0\\
  i\sin(\gamma/2)
\end{bmatrix}
\end{align*}

\subsection{A two-agent algorithm}
Following Ref. \cite{sim2011}, here we will propose a two-agent
version of algorithm that simulates the quantum operations and
measurements in Step 4-7 of $\hat{G}$ given in Section 3.2. The
entanglement measurement $\gamma$ can be simply set as its maximum
$\pi/2$. The inputs and outputs of the two-agent algorithm are shown
in Fig. 2. The \emph{Matlab} program is shown in Fig. 3(a)-(d).

\begin{figure}[!t]
\centering
\includegraphics[height=1.9in,clip,keepaspectratio]{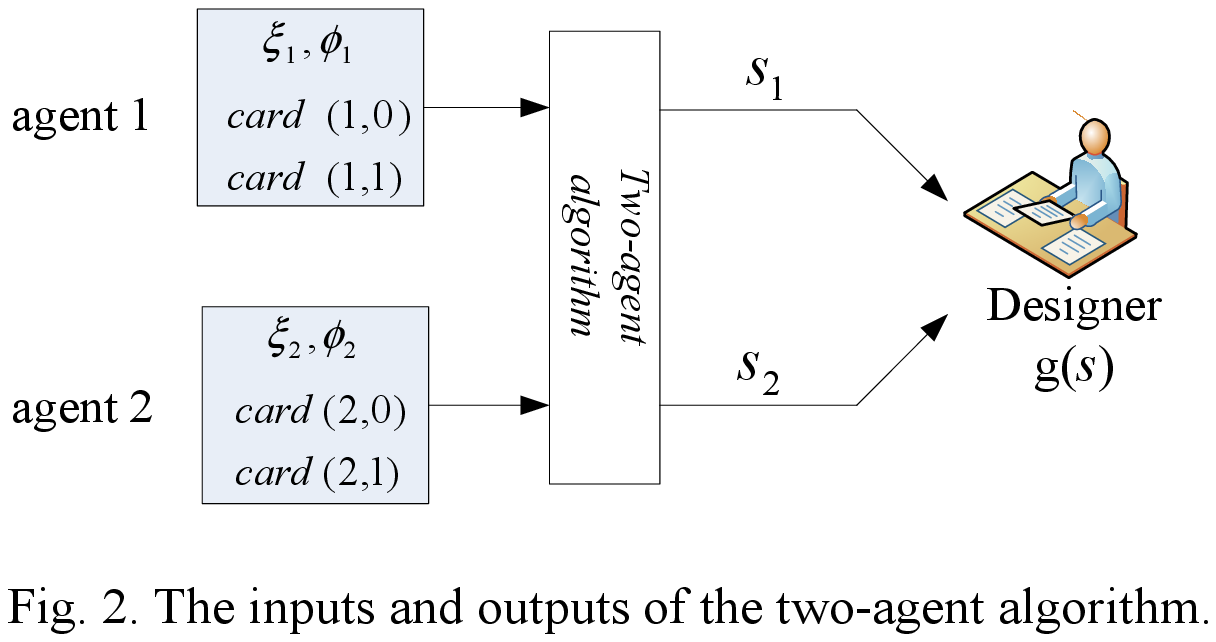}
\end{figure}

\textbf{Inputs}:\\
1) ($\xi_{j}, \phi_{j}$), $j=1, 2$: the parameters of agent $j$'s
local operation $\hat{\omega}_{j}$,
$\xi_{j}\in[0,\pi],\phi_{j}\in[0,\pi/2]$.\\
2) $card(j,0),card(j,1)\in S_{j}$, $j=1,2$: the information written
on the two sides of agent $j$'s card.

\textbf{Outputs}:\\
$s_{j}\in S_{j}$, $j=1, 2$: the strategy of agent $j$ that is sent
to the designer.

\textbf{Procedures of the algorithm}:\\
Step 1: Reading two parameters $\xi_{j}$ and $\phi_{j}$ from each
agent $j$ (See Fig. 3(a)).\\
Step 2: Computing the leftmost and rightmost columns of
$\hat{\omega}_{1}\otimes\hat{\omega}_{2}$ (See Fig. 3(b)).\\
Step 3: Computing the vector representation of
$|\psi_{2}\rangle=[\hat{\omega}_{1}\otimes\hat{\omega}_{2}]\hat{J}_{\pi/2}|CC\rangle$.\\
Step 4: Computing the vector representation of
$|\psi_{3}\rangle=\hat{J}^{+}_{\pi/2}|\psi_{2}\rangle$.\\
Step 5: Computing the probability distribution
$\langle\psi_{3}|\psi_{3}\rangle$ (See Fig. 3(c)).\\
Step 6: Randomly choosing a ``collapsed'' state from the set of all
four possible states $\{|CC\rangle, |CD\rangle, |DC\rangle,
|DD\rangle\}$ according to the probability distribution
$\langle\psi_{3}|\psi_{3}\rangle$.\\
Step 7: For each $j\in I$, the algorithm sends $card(j,0)$ (or
$card(j,1)$) as $s_{j}$ to the designer if the $j$-th basis vector
of the ``collapsed'' state is $|C\rangle$ (or $|D\rangle$) (See Fig.
3(d)).

\subsection{A two-agent version of algorithmic mechanism}
Given a two-agent algorithm that simulates the quantum operations
and measurements, the two-agent quantum mechanism
$\hat{G}:(\hat{\Omega}_{1}\otimes\hat{\Omega}_{2})\times(C_{1}\times
C_{2})\rightarrow A$ can be updated to a two-agent algorithmic
mechanism
$\widetilde{G}:(\Xi_{1}\times\Phi_{1})\times(\Xi_{2}\times\Phi_{2})\times
(C_{1}\times C_{2})\rightarrow A$, where $\Xi_{1}=\Xi_{2}=[0,\pi]$,
$\Phi_{1}=\Phi_{2}=[0,\pi/2]$.

We use $\widetilde{S}_{j}$ to express $[0,\pi]\times[0,\pi/2]\times
C_{j}$, and $\widetilde{S}_{-j}$ to express
$[0,\pi]\times[0,\pi/2]\times C_{k}$ ($k\neq j$). And thus, a
strategy profile is
$\widetilde{s}=(\widetilde{s}_{j},\widetilde{s}_{-j})$, where
$\widetilde{s}_{j}=(\xi_{j},\phi_{j},c_{j})\in\widetilde{S}_{j}$ and
$\widetilde{s}_{-j}=(\xi_{-j},\phi_{-j},c_{-j})\in\widetilde{S}_{-j}$.
A \emph{Nash equilibrium} of a two-agent algorithmic mechanism
$\widetilde{G}$ played in a preference profile $\theta$ is a
strategy profile
$\widetilde{s}^{*}=(\widetilde{s}^{*}_{1},\widetilde{s}^{*}_{2})$
such that for any agent $j\in I$,
$\widetilde{s}_{j}\in\widetilde{S}_{j}$,
$\widetilde{G}(\widetilde{s}^{*}_{1},\widetilde{s}^{*}_{2})R_{j}(\theta)
\widetilde{G}(\widetilde{s}_{j},\widetilde{s}^{*}_{-j})$.

\textbf{Working steps of the two-agent algorithmic mechanism
$\widetilde{G}$}:

Step 1: Given an SCR $f$ and a preference profile $\theta$, if the
two following conditions are satisfied, goto Step 3:\\
1) There exists $\theta'\in\Theta$, $\theta'\neq \theta$ such that
$a'R_{j}(\theta)a$ (where $a'\in f(\theta')$, $a\in f(\theta)$) for
each agent $j\in I$, and $a'P_{k}(\theta)a$ for at least one agent
$k\in I$;\\
2) If there exists $\theta''\in\Theta$, $\theta''\neq \theta'$ that
satisfies the former condition, then $a'R_{j}(\theta)a''$ (where
$a'\in f(\theta')$, $a''\in f(\theta'')$) for each agent $j\in I$,
and $a'P_{k}(\theta)a''$ for at least one agent $k\in I$.\\
Step 2: Each agent $j$ sets
$card(j,0)=(\theta_{j},a_{j},b_{j},n_{j})$ and sends $card(j,0)$ as
the strategy $s_{j}$ to the designer. Goto Step 5.\\
Step 3: Each agent $j$ sets $card(j,0)=(\theta', a',*,0)$ and
$card(j,1)=(\theta_{j},a_{j},b_{j},n_{j})$, then submits $\xi_{j}$,
$\phi_{j}$, $card(j,0)$ and $card(j,1)$ to the two-agent
algorithm.\\
Step 4: The two-agent algorithm runs in a computer and outputs
strategies $s_{1}$ and $s_{2}$ to the designer.\\
Step 5: The designer receives the overall strategy $s=(s_{1},s_{2})$
and let the final outcome be $g(s)$ using rules (1)-(6) of the
Moore-Repullo's mechanism. END.

\subsection{New result for two-agent Nash implementation}
As we have seen, in the two-agent algorithmic mechanism
$\widetilde{G}$, the entanglement measurement $\gamma$ is reduced to
its maximum $\pi/2$. Hence, condition $\lambda'$ shall be revised as
$\lambda'^{\pi/2}$, where $\lambda'^{\pi/2}_{1}$,
$\lambda'^{\pi/2}_{2}$ and $\lambda'^{\pi/2}_{3}$ are the same as
$\lambda'_{1}$, $\lambda'_{2}$ and $\lambda'_{3}$
respectively. $\lambda'^{\pi/2}_{4}$ is revised as follows:\\
$\lambda'^{\pi/2}_{4}$: For each agent $j\in I$, let him/her be the
second agent and consider his/her payoff, $\$_{CC}>\$_{DC}$.

\textbf{Proposition 2:} For two agents, given a preference profile
$\theta\in\Theta$ and an SCR $f$ that satisfies condition $\mu2$:\\
1) If condition $\lambda'^{\pi/2}$ is satisfied, then $f$ is
not Nash implementable.\\
2) If condition $\lambda'^{\pi/2}$ is not satisfied, then $f$ is
Nash implementable. Put differently, the sufficient and necessary
conditions for Nash implementation with two agents are updated as
condition $\mu2$ and no-$\lambda'^{\pi/2}$.

The proof is straightforward according to Proposition 1 in Ref.
\cite{sim2011}. Obviously, the two-agent algorithmic mechanism
proposed here is a completely ``\emph{classical}'' one that can be
run in a computer.

\section{Conclusions}
This paper generalizes the quantum and algorithmic mechanisms in
Refs. \cite{Wu2010, sim2011} to the case of two-agent Nash
implementation. Although Moore and Repullo used the phrase ``a full
characterization'' to claim that the problem of two-agent Nash
implementation had been completely solved, we argue that there
exists a new result as Proposition 2 specifies.

Since the two-agent quantum mechanism only requires two qubits to
work, theoretically current experimental technologies of quantum
information are adequate \cite{Ladd2010}. Moreover, the problem of
time and space complexity existed in the algorithmic mechanism
\cite{sim2011} does not exist here because the number of agents are
exactly two. Therefore, the two-agent algorithmic mechanism can be
applied to practical cases immediately. In this sense, the new
result on two-agent Nash implementation holds not only in the
quantum world, but also in the macro world.

\section*{Appendix}
Consider the SCR $f$ specified by Table 1. $I=\{1,2\}$,
$A=\{a^{1},a^{2},a^{3},a^{4}\}$, $\Theta=\{\theta^{1},\theta^{2}\}$.
Let $B=A$ and $C_{j}(a,\theta)=L_{j}(a,\theta)$ for each $j\in I$,
$\theta\in \Theta$, $a\in f(\theta)$, i.e.,
\begin{align*}
&C_{1}(a^{1},\theta^{1})=L_{1}(a^{1},\theta^{1})=\{a^{1},a^{2},a^{4}\},\\
&C_{2}(a^{1},\theta^{1})=L_{2}(a^{1},\theta^{1})=\{a^{1},a^{3},a^{4}\},\\
&C_{1}(a^{2},\theta^{2})=L_{1}(a^{2},\theta^{2})=\{a^{2},a^{3}\},\\
&C_{2}(a^{2},\theta^{2})=L_{2}(a^{2},\theta^{2})=\{a^{2},a^{4}\}.
\end{align*}

Obviously,
\begin{align*}
&a^{1}\in M_{1}(C_{1}(a^{1},\theta^{1}),\theta^{1})=\{a^{1}\},\\
&a^{1}\in M_{2}(C_{2}(a^{1},\theta^{1}),\theta^{1})=\{a^{1}\},\\
&a^{2}\in M_{1}(C_{1}(a^{2},\theta^{2}),\theta^{2})=\{a^{2}\},\\
&a^{2}\in M_{2}(C_{2}(a^{2},\theta^{2}),\theta^{2})=\{a^{2}\}.
\end{align*}

For each 4-tuple $(a,\theta,a',\theta')\in A\times\Theta\times
A\times\Theta$, let
\begin{align*}
&e(a^{1},\theta^{1},a^{1},\theta^{1})=a^{1}\in
C_{1}(a^{1},\theta^{1})\cap
C_{2}(a^{1},\theta^{1})=\{a^{1},a^{4}\},\\
&e(a^{1},\theta^{1},a^{2},\theta^{2})=a^{4}\in
C_{1}(a^{1},\theta^{1})\cap
C_{2}(a^{2},\theta^{2})=\{a^{2},a^{4}\},\\
&e(a^{2},\theta^{2},a^{1},\theta^{1})=a^{3}\in
C_{1}(a^{2},\theta^{2})\cap
C_{2}(a^{1},\theta^{1})=\{a^{3}\},\\
&e(a^{2},\theta^{2},a^{2},\theta^{2})=a^{2}\in
C_{1}(a^{2},\theta^{2})\cap C_{2}(a^{2},\theta^{2})=\{a^{2}\}.
\end{align*}

Case 1): Consider $\theta^{*}=\theta^{1}$,
$f(\theta^{*})=\{a^{1}\}$.

\emph{For rule (i)}:
\begin{align*}
&M_{1}(C_{1}(a^{1},\theta^{1}),\theta^{*})\cap
  M_{2}(C_{2}(a^{1},\theta^{1}),\theta^{*})=\{a^{1}\}\cap\{a^{1}\}=\{a^{1}\},\\
&M_{1}(C_{1}(a^{2},\theta^{2}),\theta^{*})\cap
  M_{2}(C_{2}(a^{2},\theta^{2}),\theta^{*})=\{a^{3}\}\cap\{a^{2}\}=\phi.
\end{align*}
Hence, rule (i) is satisfied.

\emph{For rule (ii)}:
\begin{align*}
&M_{1}(C_{1}(a^{1},\theta^{1}),\theta^{*})\cap
  M_{2}(B,\theta^{*})=\{a^{1}\}\cap\{a^{2}\}=\phi,\\
&M_{1}(C_{1}(a^{2},\theta^{2}),\theta^{*})\cap
  M_{2}(B,\theta^{*})=\{a^{3}\}\cap\{a^{2}\}=\phi,\\
&M_{2}(C_{2}(a^{1},\theta^{1}),\theta^{*})\cap
  M_{1}(B,\theta^{*})=\{a^{1}\}\cap\{a^{3}\}=\phi,\\
&M_{2}(C_{2}(a^{2},\theta^{2}),\theta^{*})\cap
  M_{1}(B,\theta^{*})=\{a^{2}\}\cap\{a^{3}\}=\phi.
\end{align*}
Hence, rule (ii) is satisfied.

\emph{For rule (iii)}:
\begin{align*}
M_{1}(B,\theta^{*})\cap
M_{2}(B,\theta^{*})=\{a^{3}\}\cap\{a^{2}\}=\phi.
\end{align*}
Hence, rule (iii) is satisfied.

\emph{For rule (iv)}:
\begin{align*}
&e(a^{1}, \theta^{1}, a^{1}, \theta^{1})=a^{1},\;
M_{1}(C_{1}(a^{1},\theta^{1}),\theta^{*})\cap
  M_{2}(C_{2}(a^{1},\theta^{1}),\theta^{*})=\{a^{1}\}\cap\{a^{1}\}=\{a^{1}\},\\
&e(a^{1}, \theta^{1}, a^{2}, \theta^{2})=a^{4},\;
M_{1}(C_{1}(a^{1},\theta^{1}),\theta^{*})\cap
  M_{2}(C_{2}(a^{2},\theta^{2}),\theta^{*})=\{a^{1}\}\cap\{a^{2}\}=\phi,\\
&e(a^{2}, \theta^{2}, a^{1}, \theta^{1})=a^{3},\;
M_{1}(C_{1}(a^{2},\theta^{2}),\theta^{*})\cap
  M_{2}(C_{2}(a^{1},\theta^{1}),\theta^{*})=\{a^{3}\}\cap\{a^{1}\}=\phi,\\
&e(a^{2}, \theta^{2}, a^{2}, \theta^{2})=a^{2},\;
M_{1}(C_{1}(a^{2},\theta^{2}),\theta^{*})\cap
  M_{2}(C_{2}(a^{2},\theta^{2}),\theta^{*})=\{a^{3}\}\cap\{a^{2}\}=\phi.
\end{align*}
Hence, rule (iv) is satisfied.

Case 2): Consider $\theta^{*}=\theta^{2}$,
$f(\theta^{*})=\{a^{2}\}$.

\emph{For rule (i)}:
\begin{align*}
&M_{1}(C_{1}(a^{1},\theta^{1}),\theta^{*})\cap
  M_{2}(C_{2}(a^{1},\theta^{1}),\theta^{*})=\{a^{4}\}\cap\{a^{3}\}=\phi,\\
&M_{1}(C_{1}(a^{2},\theta^{2}),\theta^{*})\cap
  M_{2}(C_{2}(a^{2},\theta^{2}),\theta^{*})=\{a^{2}\}\cap\{a^{2}\}=\{a^{2}\}.
\end{align*}
Hence, rule (i) is satisfied.

\emph{For rule (ii)}:
\begin{align*}
&M_{1}(C_{1}(a^{1},\theta^{1}),\theta^{*})\cap
  M_{2}(B,\theta^{*})=\{a^{4}\}\cap\{a^{3}\}=\phi,\\
&M_{1}(C_{1}(a^{2},\theta^{2}),\theta^{*})\cap
  M_{2}(B,\theta^{*})=\{a^{2}\}\cap\{a^{3}\}=\phi,\\
&M_{2}(C_{2}(a^{1},\theta^{1}),\theta^{*})\cap
  M_{1}(B,\theta^{*})=\{a^{3}\}\cap\{a^{4}\}=\phi,\\
&M_{2}(C_{2}(a^{2},\theta^{2}),\theta^{*})\cap
  M_{1}(B,\theta^{*})=\{a^{2}\}\cap\{a^{4}\}=\phi.
\end{align*}
Hence, rule (ii) is satisfied.

\emph{For rule (iii)}:
\begin{align*}
M_{1}(B,\theta^{*})\cap
M_{2}(B,\theta^{*})=\{a^{4}\}\cap\{a^{3}\}=\phi.
\end{align*}
Hence, rule (iii) is satisfied.

\emph{For rule (iv)}:
\begin{align*}
&e(a^{1}, \theta^{1}, a^{1}, \theta^{1})=a^{1},\;
M_{1}(C_{1}(a^{1},\theta^{1}),\theta^{*})\cap
  M_{2}(C_{2}(a^{1},\theta^{1}),\theta^{*})=\{a^{4}\}\cap\{a^{3}\}=\phi,\\
&e(a^{1}, \theta^{1}, a^{2}, \theta^{2})=a^{4},\;
M_{1}(C_{1}(a^{1},\theta^{1}),\theta^{*})\cap
  M_{2}(C_{2}(a^{2},\theta^{2}),\theta^{*})=\{a^{4}\}\cap\{a^{2}\}=\phi,\\
&e(a^{2}, \theta^{2}, a^{1}, \theta^{1})=a^{3},\;
M_{1}(C_{1}(a^{2},\theta^{2}),\theta^{*})\cap
  M_{2}(C_{2}(a^{1},\theta^{1}),\theta^{*})=\{a^{2}\}\cap\{a^{3}\}=\phi,\\
&e(a^{2}, \theta^{2}, a^{2}, \theta^{2})=a^{2},\;
M_{1}(C_{1}(a^{2},\theta^{2}),\theta^{*})\cap
  M_{2}(C_{2}(a^{2},\theta^{2}),\theta^{*})=\{a^{2}\}\cap\{a^{2}\}=\{a^{2}\}.
\end{align*}
Hence, rule (iv) is satisfied.

To sum up, the SCR $f$ given in Table 1 satisfies condition $\mu2$.
Therefore, according to Moore-Repullo's theorem, it \emph{should be}
Nash implementable. However, as shown in Section 3 and 4, when
condition $\lambda'$ is satisfied, neither in the quantum world nor
in the macro world will the SCR $f$ be Nash implementable.

\section*{Acknowledgments}

The author is very grateful to Ms. Fang Chen, Hanyue Wu
(\emph{Apple}), Hanxing Wu (\emph{Lily}) and Hanchen Wu
(\emph{Cindy}) for their great support.


\newpage
\begin{figure}[!t]
\centering
\includegraphics[height=2.5in,clip,keepaspectratio]{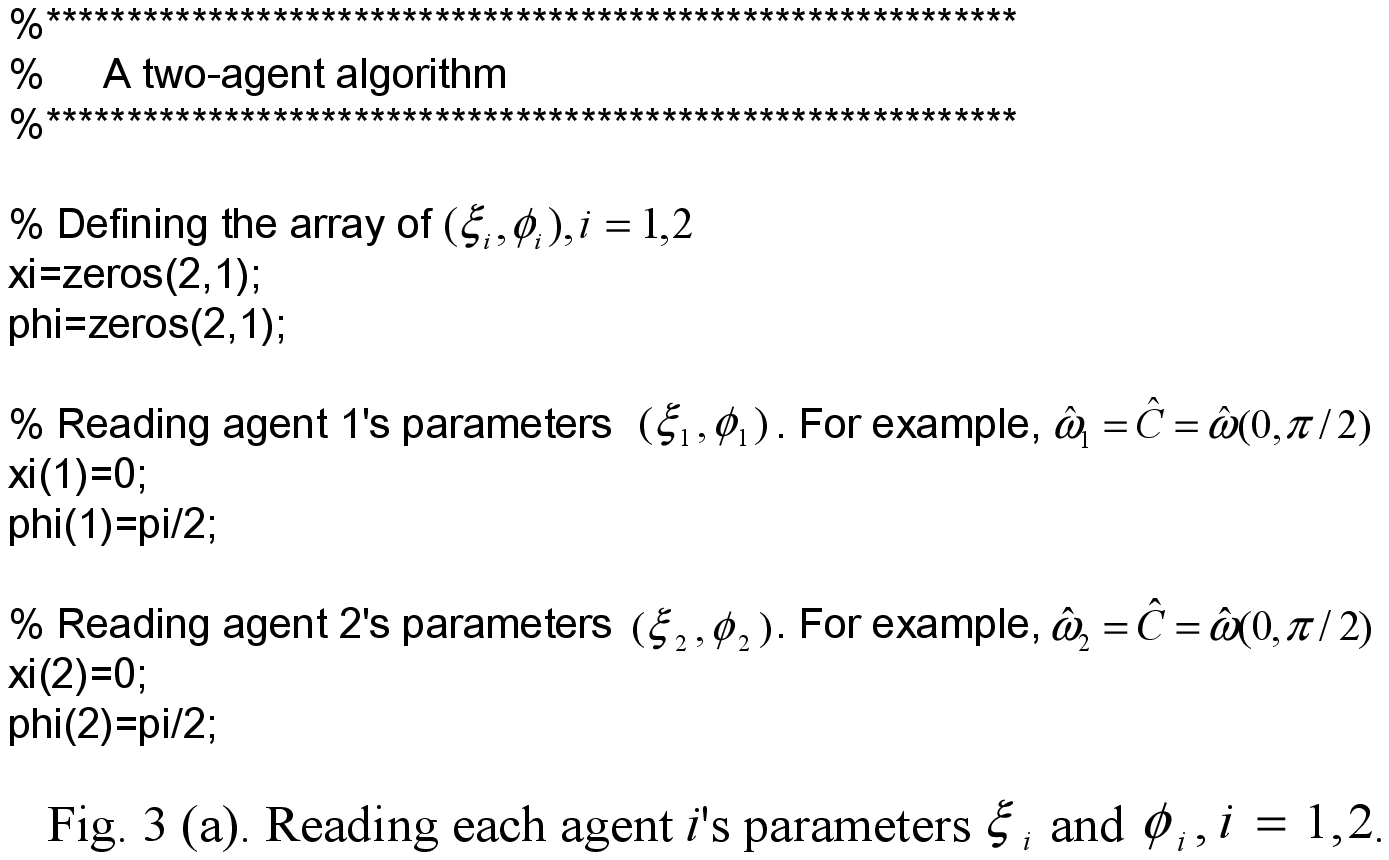}
\end{figure}

\begin{figure}[!t]
\centering
\includegraphics[height=4.3in,clip,keepaspectratio]{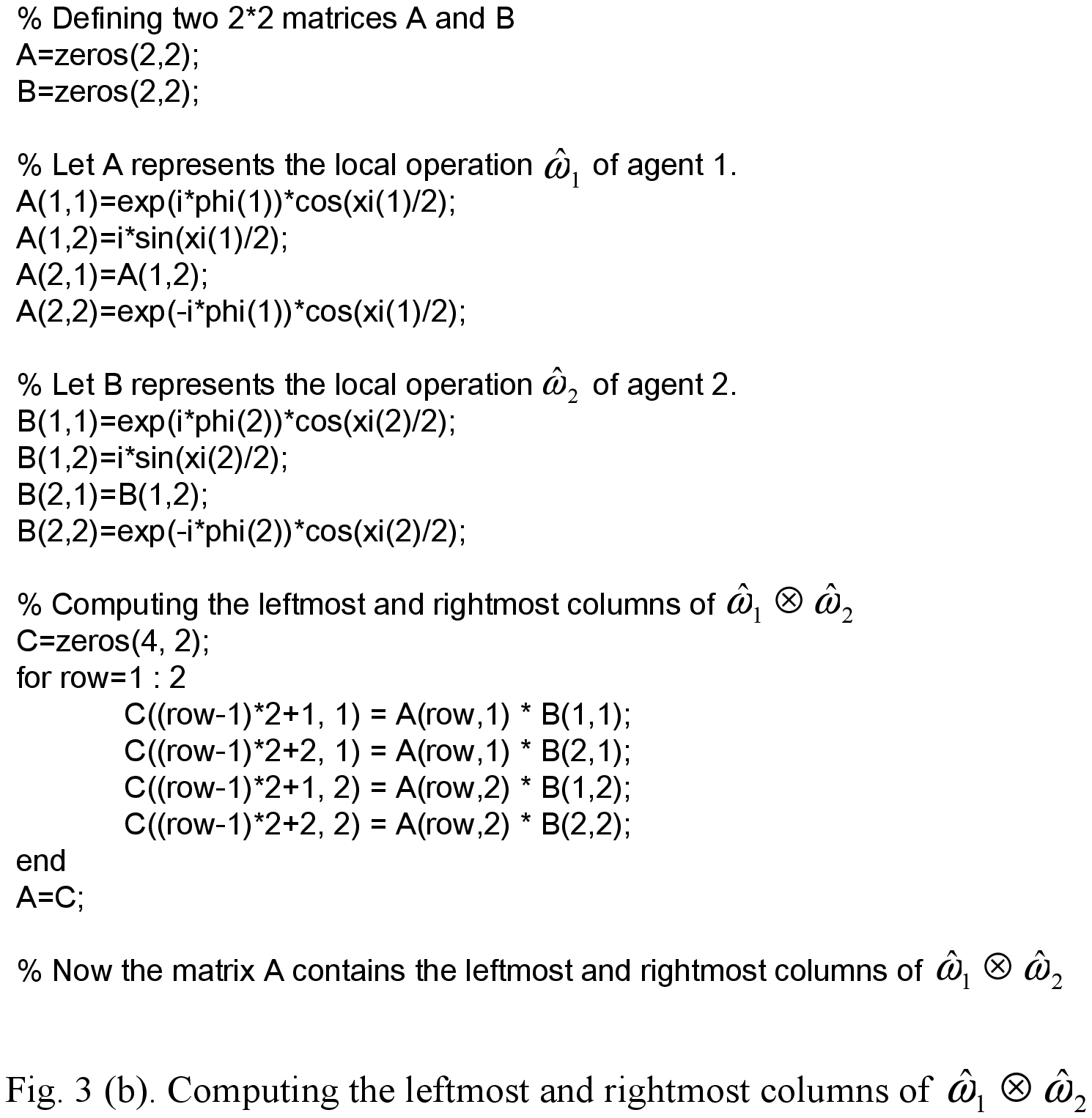}
\end{figure}

\newpage
\begin{figure}[!t]
\centering
\includegraphics[height=2.6in,clip,keepaspectratio]{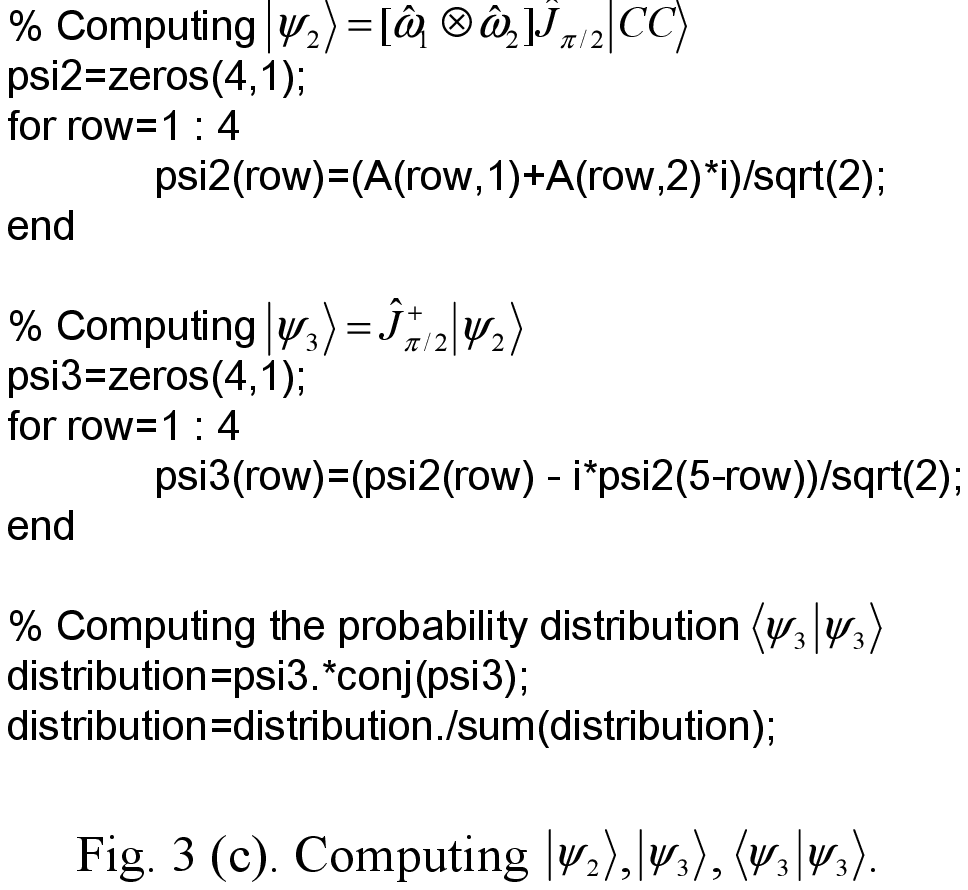}
\end{figure}

\begin{figure}[!t]
\centering
\includegraphics[height=5.5in,clip,keepaspectratio]{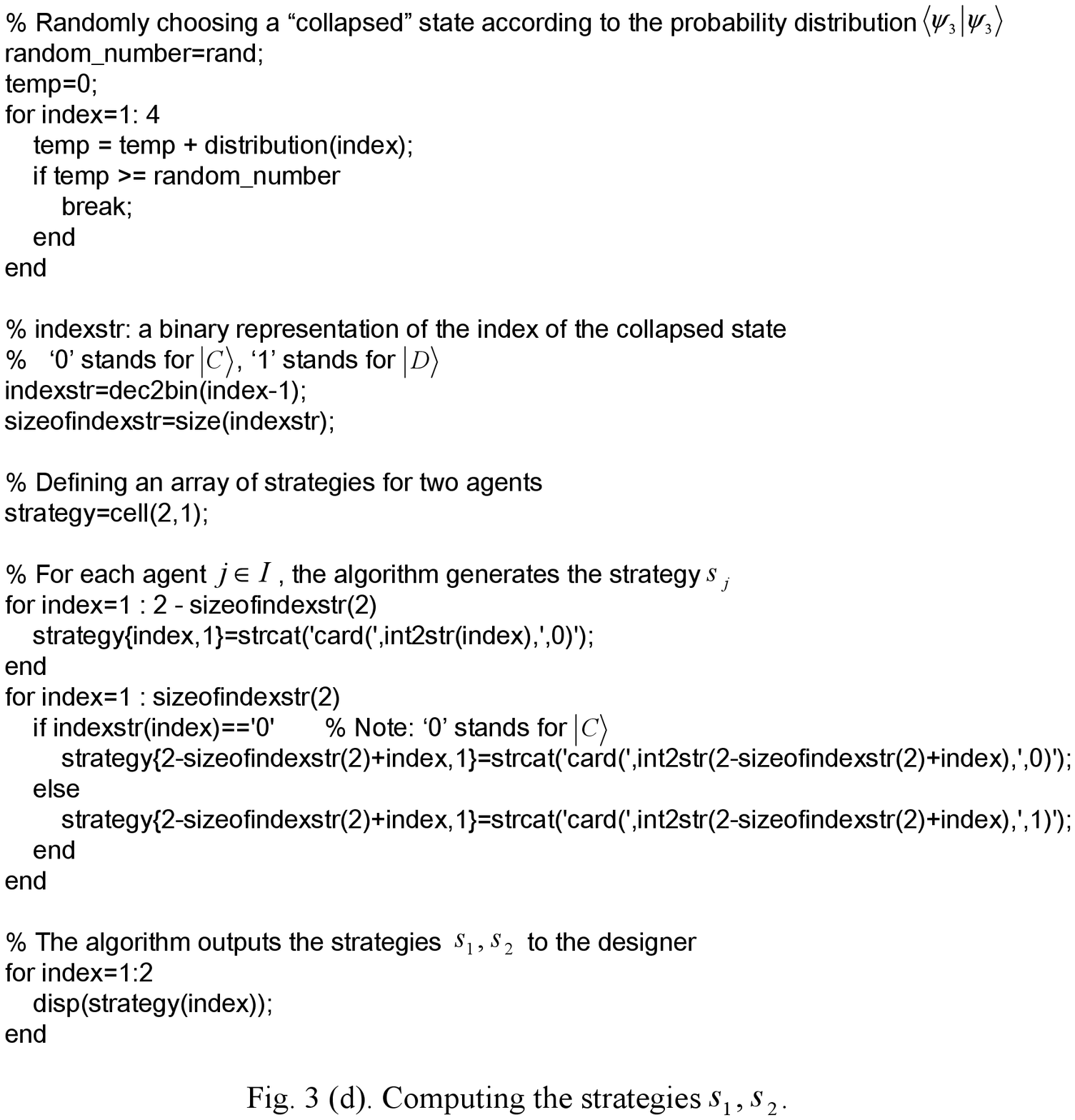}
\end{figure}

\end{document}